\documentclass[9pt,twocolumn,twoside]{opticajnl}
\journal{opticajournal} 

\setboolean{shortarticle}{true}
\newcommand{\vd}[0]{\textit{vanishing distance~}}

\usepackage{lineno}
\usepackage{makecell}

\title{The \textit{vanishing distance}: a depth-range boundary for dynamic wavefront shaping}

\date{23 Sept. 2026}

\author[1, +]{Hugo Lassiette}
\author[1, +]{Léa Testé}
\author[1]{Léa Krafft} 
\author[1]{Geoffrey Maulion}
\author[1]{Vincent Michau}
\author[2]{Romain Pierrat} 
\author[2]{Sébastien Popoff}
\author[2,3,4]{Willem L. Vos}
\author[1,2,*]{Serge Meimon}
\affil[1]{DOTA, ONERA, Université Paris-Saclay, 92320 Châtillon, France}
\affil[2]{Institut Langevin, ESPCI Paris, PSL University, CNRS, 1 rue Jussieu, 75005 Paris, France}
\affil[3]{Complex Photonic Systems (COPS) Chair, Faculty of Science and Technology, University of Twente, 7500 AE Enschede, The Netherlands}
\affil[4]{Complex Photonic Systems (COPS) group, Photonic and Semiconductor Nanostructures (PSN) Chair, Department of Applied Physics and Science Education (APSE), Eindhoven University of Technology (TU/e), P.O. Box 513, 5600 MB Eindhoven, the Netherlands}
\affil[+]{These authors contributed equally.}
\affil[*]{Corresponding author: \href{mailto:serge.meimon@onera.fr}{serge.meimon@onera.fr}}

\begin{abstract}
Optical imaging and transmission through scattering media conventionally rely on ballistic light, whose power decays exponentially with the optical thickness. 
{Although ballistic light cannot be exploited beyond a certain depth,} no simple criterion currently identifies this limit. 
{Therefore, we introduce here} the \vd {that is} defined as the propagation length at which the ballistic power equals the power carried by a single diffuse optical mode. 
Beyond {the vanishing} distance, multiply scattered light dominates, which can be harnessed 
by optical wavefront shaping or by computational reconstructions. 
We present a simple semi-analytical model for radiance propagation, validated {by} Monte Carlo simulations and {by} experiments in both monodisperse and heterogeneous scattering media. 
{From these data, we estimate} the \vd and the associated minimum source power. 
Across representative scattering environments, ballistic light vanishes after about 20 {to} 40 scattering mean free paths. 
Unlike the transport mean free path that characterizes the angular randomization of the scattered {light}, the \vd identifies the loss of the ballistic channel on a modal-power basis. 
The \vd {is thus} 
a two-sided bound: ballistic approaches cannot be efficiently applied beyond it, whereas wavefront shaping of {scattered} 
light offers no advantage below it. 
\end{abstract}

\setboolean{displaycopyright}{false} 

\begin{document}

\maketitle

Optical imaging and transmission through scattering media conventionally rely on ballistic light, the fraction of the incident light that has not been scattered and therefore retains the spatial information of the source; confocal, multiphoton, and coherence-gated microscopies, as well as free-space optical links, preferentially select this component through spatial, temporal, or coherence gating~\cite{Dunsby2003JPD, Ntziachristos2010NatMet, badon2016smart}. 
However, the ballistic power decays exponentially with the optical thickness, following the Bouguer--Beer--Lambert law, until it eventually drops below the multiply scattered background. 
This sets the range of ballistic techniques, and the question arises where this limit lies, {that is, beyond which distance can} ballistic light no longer be exploited? 

Beyond that {distance}, the transmitted light is dominated by multiply scattered and diffuse light. 
Although it forms a seemingly random speckle, {multiple scattered} light remains coherent and deterministic, and can still be exploited if its input wavefront is controlled or its link to relevant parameters is estimated. 
Wavefront shaping (WFS) is one of the most popular approaches to do so, and has led to major advances in imaging, communication, and beam delivery through scattering media~\cite{Vellekoop2007OL, Mosk2012NP, Gigan2022JPhys, Cao2022NatPhys, balasiano-2024}. 
In static or slowly evolving samples, digital optical phase conjugation and transmission-matrix approaches have enabled diffraction-limited focusing, deep imaging, and efficient light delivery through highly scattering materials~\cite{Aubry2009AO, Popoff2011PRL, Badon2020SA, Feldkhun2019Optica}. 

However, the relevance of WFS {to both} dynamic and extended scattering media, such as fog, seawater, or biological flows, remains limited~\cite{Pine1990JPhys, Ntziachristos2010NatMet, Mididoddi2025NatPho}.
In these environments, rapid temporal decorrelations, large optical thicknesses, and {considerable photon limitations severely} restrict the conditions under which coherent wavefront control can be
implemented~\cite{Feng1988PRL}. 
While dynamic WFS has been demonstrated in thin or weakly scattering samples~\cite{Vellekoop2007OL, Mididoddi2025NatPho}, long-{distance} applications {involving} atmospheric or underwater propagation still lack simple physical criteria that define when WFS becomes necessary and when it remains feasible. 

{It is a} common misconception to associate the transition between the regime where ballistic light remains usable and the regime where one resorts to WFS with the transport mean free path $\ell_t$.
In reality, $\ell_t$ {does} characterize the progressive loss of the directional memory of multiply scattered light and provides the characteristic scale over which diffuse-radiance approximations become applicable \cite{vanRossum1999RMP,Rotter2017RMP}. 
{But} $\ell_t$ does not indicate the disappearance of the ballistic component, which may remain detectable over several transport mean free paths, depending on the scattering anisotropy and observation geometry. 
Consequently, $\ell_t$ does not determine when ballistic approaches cease to be applicable. 

{Therefore, we introduce here} the \emph{vanishing distance} $\ell_v$ that is defined as the propagation distance {where} the ballistic power equals the power carried by a single diffuse optical mode. 
Beyond this distance, the ballistic contribution is no longer distinguishable from the scattered background on a modal basis, and transmission or imaging can no longer rely on the ballistic component. 
The \vd therefore marks the onset of the regime in which further propagation requires the exploitation of the {scattered fields}. 
This bound is derived under idealized conditions that we make explicit. 
We assume that {in the ballistic regime the} light remains diffraction limited, i.e., concentrated in a single optical mode, either because the medium introduces no aberration or because adaptive optics ideally corrects it~\cite{roddier1999adaptive, krafft2024adaptive, Aubry2009AO, Badon2020SA}. 
Below $\ell_v$, sufficient ballistic photons are collected in this single mode, and shaping the {scattered} light is not necessary. 
In the diffuse, {scattering regime} beyond $\ell_v$, the transmitted power is spread over many {scattering} modes, and one must resort to WFS to harness {the light}, as in the examples discussed above. 
In this sense, $\ell_v$ is a limit rather than a guarantee, much like a Cramér--Rao bound: 
{whereas a certain} practical implementation may lose the ballistic channel before $\ell_v$ or fail to exploit the {scattered} light beyond $\ell_v$, ballistic approaches \emph{cannot} {effectively operate} beyond $\ell_v$, and WFS of {scattered light offers} \emph{no advantage} below $\ell_v$. 

To evaluate $\ell_v$, we develop a simple semi-analytical model of diffuse radiance that combines analytical asymptotic expressions with Monte Carlo simulations. 
The model is validated experimentally using both {suspension of monodisperse polystyrene nanospheres as well as} a heterogeneous medium of polydisperse non-spherical sugar microparticles {suspended in oil}. 
Finally, we introduce the associated vanishing power $P_v$, defined as the minimum source power required to reach the single-photon-per-mode threshold at $\ell_v$ within the available integration time~\cite{Paterson2008}. 
We evaluate both $\ell_v$ and $P_v$ for representative scattering media and show that the ballistic light typically vanishes in the diffuse background after approximately 20--40 scattering mean free paths.

\begin{figure}[ht]
\centering
\includegraphics[width=0.9\linewidth]{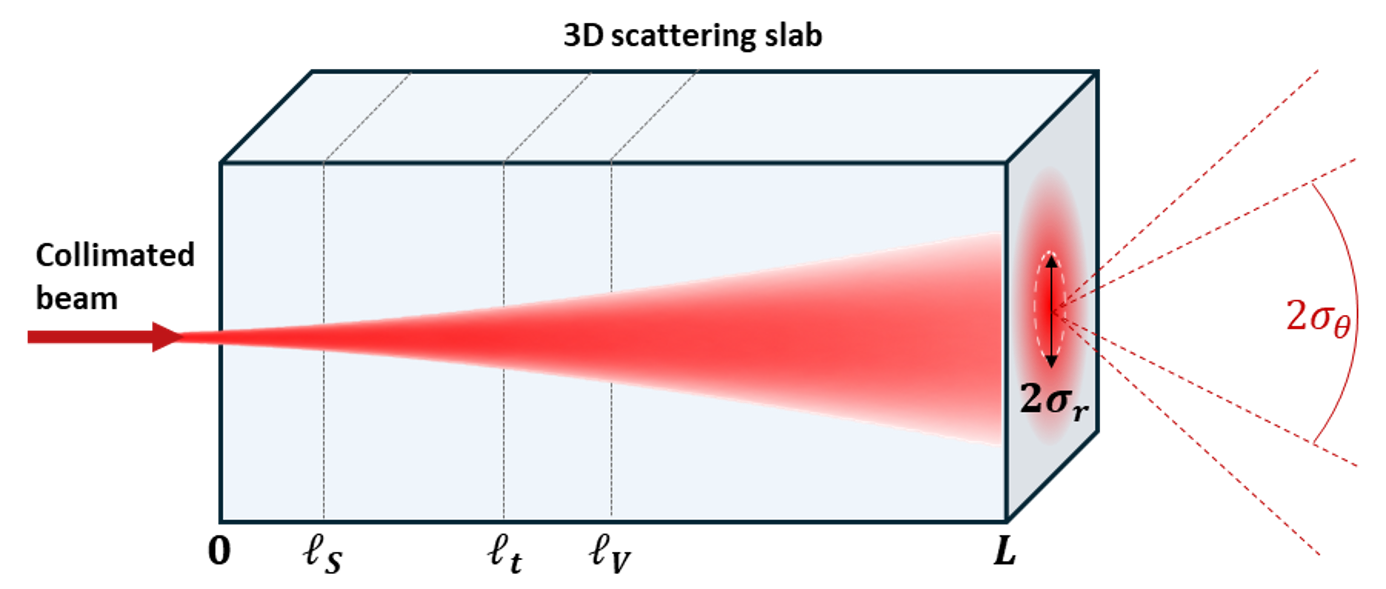}
\caption{Geometry considered throughout the paper. A narrow collimated beam illuminates a three-dimensional scattering slab of thickness $L$. The transmitted scattered field forms a halo at the output plane, characterized by its radial spatial spread $\sigma_r$ and angular spread $\sigma_\theta$. The \vd $\ell_v$ marks the modal-power crossover between the residual ballistic component and a single diffuse optical mode.}
\label{fig:geometry}
\end{figure}

We consider a three-dimensional scattering slab of thickness $L$, illuminated at normal incidence by a narrow collimated beam whose transverse extent is small compared with $L$, as illustrated in Fig.~\ref{fig:geometry}. The incident power is denoted by $P_0$. The non-absorbing medium is characterized by its scattering mean free path $\ell_s$ and a Henyey–Greenstein phase function \cite{Modest2003} with anisotropy factor $g=\langle\cos\theta\rangle$, giving a transport mean free path $\ell_t=\ell_s/(1-g)$. The transmitted field is evaluated at the output plane of the slab.

Wavefront shaping becomes relevant when the ballistic component can no longer be exploited as a distinct optical channel. We therefore define the \emph{vanishing distance} $\ell_v$, as the propagation distance at which the residual ballistic power $P_b$ equals the power carried by a single diffuse optical mode $P_m$,
\begin{equation}
P_b(\ell_v)=P_m(\ell_v).
\label{eq:lv_definition}
\end{equation}

This definition assumes an ideal ballistic beam that remains diffraction limited and concentrated into a single optical mode throughout propagation. In practice, residual divergence, aberrations, or imperfect mode matching may distribute the ballistic energy over several modes, causing the operational loss of the ballistic channel to occur at a shorter distance. Such system-dependent effects are outside the scope of the present work. Under the ideal-mode assumption adopted here, $\ell_v$ therefore depends only on the scattering properties of the medium.

Unlike the transport mean free path $\ell_t$, which characterizes the progressive loss of the directional memory of multiply scattered light and the approach to the diffusion regime, $\ell_v$ compares the modal powers of the ballistic and diffuse components. It identifies the propagation distance beyond which ballistic imaging or transmission can no longer rely on a distinguishable ballistic contribution.

The residual ballistic power follows the Bouguer--Beer--Lambert law. At the exit surface, the power equals:
\begin{equation}
P_b(L)=P_0\exp\!\left(-\frac{L}{\ell_s}\right),
\label{eq:ballistic}
\end{equation}
where the extinction length equals $\ell_s$ under the non-absorbing assumption used throughout this Letter.

The average power $P_m$ carried by a single diffraction-limited diffuse optical mode is directly proportional to the ensemble averaged diffuse radiance at the output plane. Since one spatial--angular mode occupies an optical etendue of order $\lambda^2$ \cite{Mertz},
\begin{equation}
P_m(L)=\lambda^2\,\mathcal{L}_d(L),
\label{eq:pm}
\end{equation}
where $\lambda$ is the optical wavelength and $\mathcal{L}_d$ denotes the total transmitted diffuse radiance.
The diffuse radiance is estimated from the total diffuse transmission $T(L)$ and the spatial--angular etendue $G_s(L)$ occupied by the transmitted scattered field,
\begin{equation}
\mathcal{L}_d(L)
=
P_0(1-e^{-L/\ell_s})
\frac{T(L)}
{G_s(L)}.
\label{eq:luminance}
\end{equation}
We characterize $G_s$ through the RMS radial spatial spread $\sigma_r$ at the output plane and the RMS polar angular spread $\sigma_\theta$ of the transmitted directions.

In a dynamic medium, the available integration time is limited by the speckle
correlation time $\tau_\mathrm{corr}$. Under a single-scattering estimate, a scatterer displacement
$\Delta\mathbf{r}$ produces a phase shift
$\Delta\phi=\mathbf{q}\cdot\Delta\mathbf{r}$, where $\mathbf{q}$ is the
scattering vector \cite{joo2010diffusive}. For directional scatterer motion at a characteristic velocity $v$, taking $q\approx2\pi/\lambda$ and assuming a characteristic dephasing of $\Delta\phi=1$ rad yields the order-of-magnitude estimate
\begin{equation}
\tau_{\mathrm{corr}}\sim\frac{\lambda}{2\pi v}.
\label{eq:corr_time}
\end{equation}
This estimate also applies to a collective translation of the scattering
medium. Velocities of order $0.1~\mathrm{m\,s^{-1}}$ at near-infrared
wavelengths correspond to correlation times of order $1~\mu\mathrm{s}$.
We use $\tau_{\mathrm{corr}}=1~\mu\mathrm{s}$ in the numerical examples below.

Following the fundamental-information-limit argument of Paterson \cite{Paterson2008}, we define the associated \emph{vanishing power} $P_v$ as the incident power required to provide one photon in a single diffuse optical mode during $\tau_{\mathrm{corr}}$ at $L=\ell_v$. Using~\eqref{eq:pm} and \eqref{eq:luminance} to match this threshold,

\begin{equation*}
P_m(\ell_v)
=
\frac{hc}{\lambda\tau_{\mathrm{corr}}}=
\lambda^2P_v(1-e^{-L/\ell_s})\frac{T(\ell_v) }
{G_s(\ell_v)}.
\end{equation*}

\begin{equation}
\Rightarrow P_v
=
\frac{hc}{\lambda\tau_{\mathrm{corr}}}
\frac{G_s(\ell_v)}
{\lambda^2 (1-e^{-L/\ell_s})T(\ell_v)}.
\label{eq:vanishing_power}
\end{equation}
Thus, $P_v\propto 1/\tau_{\mathrm{corr}}$, whereas $\ell_v$ is independent of the temporal dynamics under the assumptions of the present model.

Determining the \vd and vanishing power requires the evaluation of the total diffuse transmission $T$ and the spatial--angular etendue $G_s$ throughout the transition from forward scattering to diffusive transport. We characterize the transmitted halo by its RMS radial spatial spread $\sigma_r$  and RMS polar angular spread $\sigma_\theta$ at the output plane. Approximating the radial distributions to be Gaussian, the corresponding spatial--angular etendue is 
\begin{equation}
G_s(L,\ell_s,g)
=
2\pi\sigma_r^2\,
2\pi\left[
1-\cos\left(\sqrt{2}\sigma_\theta\right)
\right].
\label{eq:scattered_etendue}
\end{equation}
Together with the total diffuse transmission $T$, these quantities determine the transmitted diffuse radiance through ~\eqref{eq:luminance}. We evaluate $T, \sigma_r$ and $\sigma_\theta$ using a hybrid radiative-transfer model combining Monte Carlo simulations at moderate optical thicknesses with analytical asymptotic expressions in the diffusive regime. Details of the model, its numerical validation, lookup tables, and Monte Carlo implementation are provided in Supplementary Sections~S2-S4.

We experimentally validate the predicted evolution of the ballistic and diffuse modal powers using three scattering phantoms. Two reference samples consist of aqueous suspensions of monodisperse polystyrene microspheres with diameters of $1~\mu\mathrm{m}$ and $3~\mu\mathrm{m}$. The third sample consists of sugar particles dispersed in soybean oil, providing a heterogeneous medium with non-spherical particles and a broad size distribution.

A collimated beam illuminates the sample, and the transmitted field is recorded on a camera. The ballistic contribution is estimated from the central peak of the transmitted intensity distribution, whereas the scattered modal power is obtained from the mean diffuse background after removal of the ballistic contribution. For each sample, the optical thickness is varied by changing the particle concentration. Experimental intensities are normalized by the incident power and acquisition time.

\begin{figure}[ht]
\centering
\includegraphics[width=\linewidth]{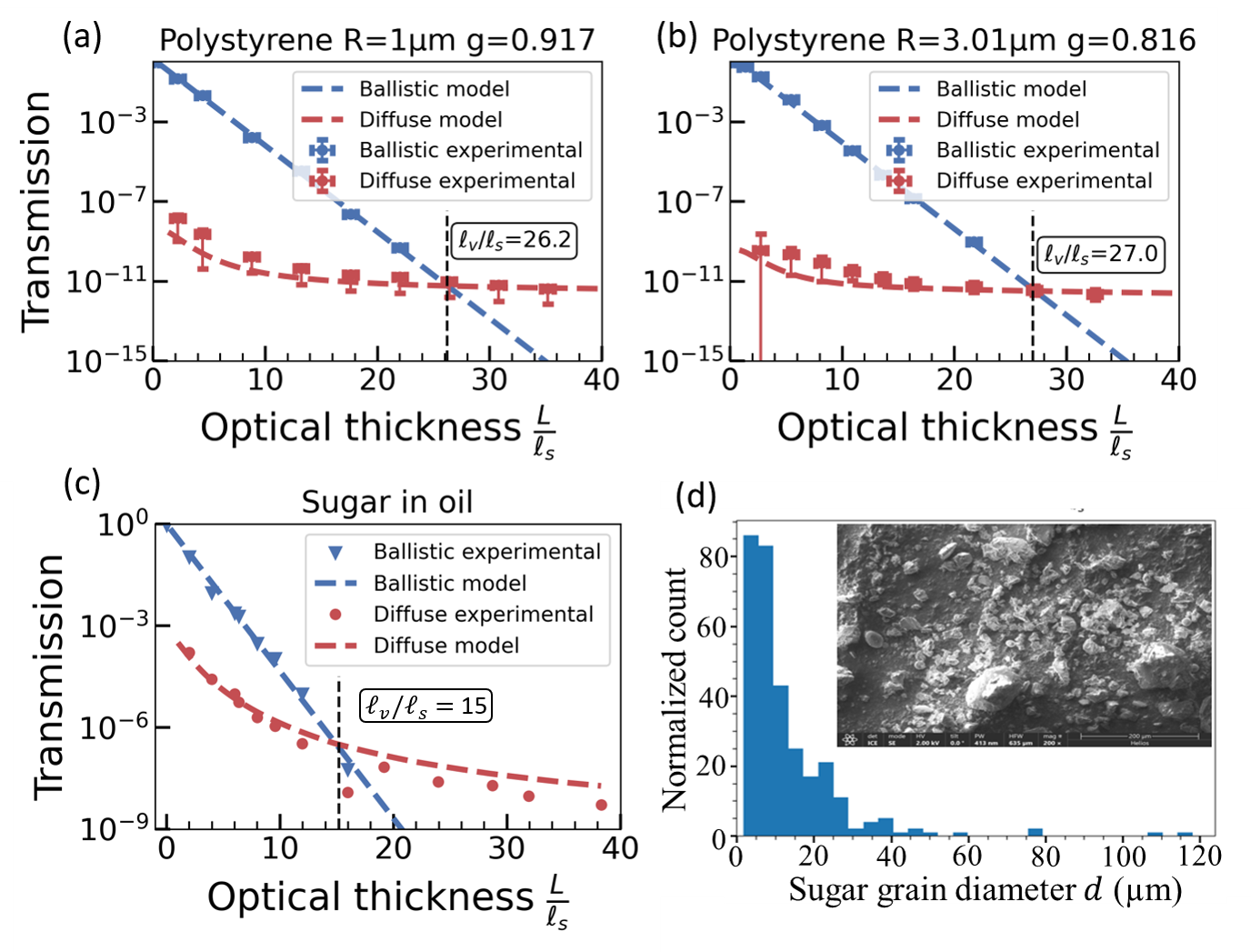}
\caption{Experimental validation of the ballistic and diffuse modal-power model. Measurements are shown for aqueous suspensions of monodisperse polystyrene microspheres with diameters of (a) $1~\mu\mathrm{m}$ and (b) $3~\mu\mathrm{m}$, and for (c) a heterogeneous suspension of sugar particles in soybean oil. Symbols denote measurements and solid lines denote model predictions. The vertical marker indicates the \vd $\ell_v$, defined by equality of the ballistic power and the power carried by one diffuse mode. (d) Particle-size distribution and scanning-electron-microscopy image of the sugar particles, showing their broad size distribution and irregular morphology.}
\label{fig:exp}
\end{figure}

Figure~\ref{fig:exp} compares the measured ballistic and scattered modal powers with the predictions of the hybrid model. For the polystyrene suspensions, the anisotropy factors are taken from Mie theory \cite{Mie1908} and the literature \cite{dule-2014}. The model reproduces both the exponential decay of the ballistic component and the slower evolution of the scattered modal power over the explored optical-thickness range.

The sugar suspension provides a more stringent test because its particles are neither spherical nor monodisperse. Their diameters extend approximately from $1$ to $50~\mu\mathrm{m}$, as shown by the size histogram and scanning-electron-microscopy image in Fig.~\ref{fig:exp}(d). Despite this heterogeneity, the measured power curves are well reproduced using a single effective anisotropy factor, $g=0.99$. This agreement suggests that an effective anisotropy factor can capture the
modal-power crossover even for a heterogeneous distribution of non-spherical
particles.

For all three samples, the crossing between the ballistic and scattered modal powers is captured by the model. The experiments therefore validate the use of the hybrid transport model to determine $\ell_v$ in both controlled monodisperse media and more realistic heterogeneous scattering environments.

\begin{figure}[ht]
\centering
\includegraphics[width=0.9\linewidth]{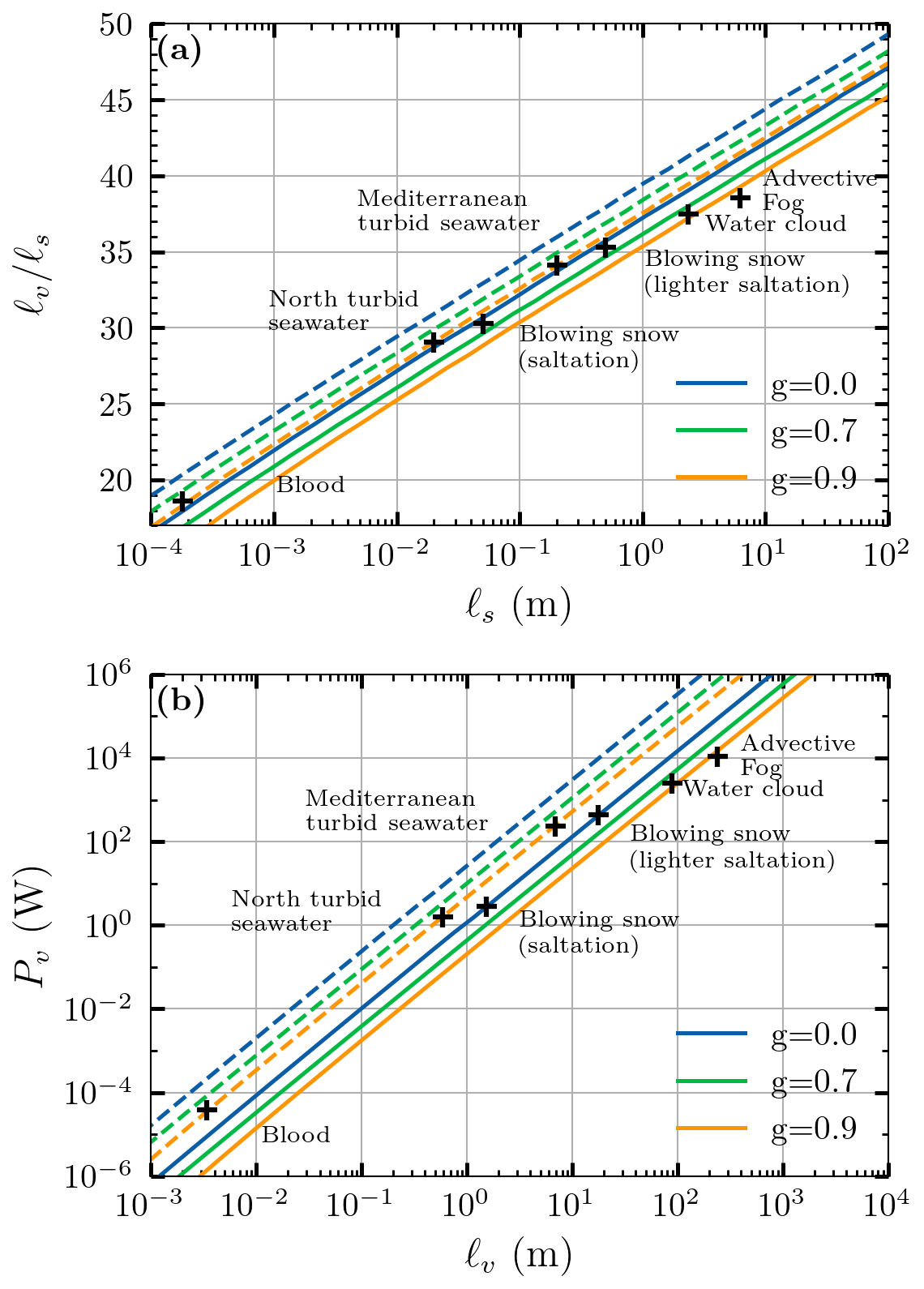}
\caption{(a) Normalized \vd $\ell_v/\ell_s$ as a function of the scattering mean free path $\ell_s$ for representative anisotropy factors. Full line: $\lambda=1.55~\mu m$. Dashed line: $\lambda=555~nm$. Markers identify the representative scattering environments listed in Supplementary Section S6. (b) Associated vanishing power $P_v$, evaluated for an integration time $\tau_{\mathrm{corr}}=1~\mu\mathrm{s}$ under the non-absorbing assumption.}
\label{fig:Pv_lv_g}
\end{figure}

The validated hybrid model can now be used to determine the \vd and the associated source power for a broad range of non-absorbing scattering slabs. Figure~\ref{fig:Pv_lv_g}(a) shows the normalized \vd $\ell_v/\ell_s$ as a function of the scattering mean free path $\ell_s$ for representative anisotropy factors and wavelengths.

Although the transport mean free path varies strongly with the anisotropy factor according to $\ell_t=\frac{\ell_s}{1-g}$,
the normalized \vd is remarkably stable. Over the six-decade range of scattering mean free path, from $\ell_s=10^{-4}$ to $10^{2}$~m, and for the anisotropy factors shown in Fig.~\ref{fig:Pv_lv_g}, $\ell_v/\ell_s$ remains confined to a narrow interval between approximately 20 and 40. The dependence on $g$ therefore remains moderate despite its strong influence on $\ell_t$. This observation provides the practical rule of thumb
\begin{equation}
\ell_v \simeq 20\text{--}40\,\ell_s,
\label{eq:lv_rule}
\end{equation}
for the range of parameters considered here.

This result emphasizes the distinct physical meanings of $\ell_t$ and $\ell_v$. The transport mean free path characterizes the progressive randomization of photon directions and the approach to the diffusive regime. In contrast, the \vd identifies the disappearance of the ballistic channel on a modal basis. These two characteristic lengths therefore describe different physical transitions and should not be interpreted interchangeably.

Using~\eqref{eq:vanishing_power}, we evaluate the source power required to reach the \vd within one speckle correlation time. Figure~\ref{fig:Pv_lv_g}(b) shows the resulting vanishing power $P_v$ for $\tau_{\mathrm{corr}}=1~\mu\mathrm{s}$. Unlike $\ell_v$, which depends only on the scattering properties under the ideal-mode assumption adopted here, $P_v$ additionally depends on the medium dynamics through $P_v\propto1/\tau_{\mathrm{corr}}$.

Representative scattering environments are indicated by markers in Fig.~\ref{fig:Pv_lv_g}. Their wavelengths, scattering mean free paths, and anisotropy factors are listed in Supplementary Section S6 \cite{Babin2003LiOc,raisanen-2015,Pomeroy1988JoG,kokhanovsky-2003,tam-1979}. These media are evaluated here using their scattering parameters only; absorption is neglected.

In summary, we have introduced the \vd, defined as the propagation distance at which the residual ballistic power equals the power carried by a single diffuse optical mode. Unlike the transport mean free path, which characterizes the angular randomization of scattered light, the \vd quantifies the modal disappearance of the ballistic channel.

Using a hybrid model combining analytical diffusive asymptotes, Monte Carlo simulations, and experimental validation, we showed that, within the non-absorbing slab geometry considered here, the normalized \vd remains remarkably stable, typically between 20 and 40 scattering mean free paths over a broad range of scattering conditions.

Combined with the associated vanishing power, this criterion provides simple metrics for estimating both the propagation range and the photometric requirements of dynamic wavefront shaping. We expect these quantities to facilitate the evaluation and comparison of wavefront-shaping strategies in representative scattering environments.
\begin{backmatter}
\bmsection{Funding} This work was supported by the French National Research Agency (ANR) under grant ANR-22-ASTR-0020 (AQUAMAT).

\bmsection{Acknowledgment} We thank Melissa Goodwin for helping characterize the sugar scatterers. We thank Ivo Vellekoop and Allard Mosk for stimulating discussions. 
We thank Ad Lagendijk for encouragement. This work also benefited from the scientific environment of the ONERA chair ``Seeing Through Fog'' (\url{https://w3.onera.fr/chaire/seethrufog/}).

WLV acknowledges support by the "Nederlandse Organisatie voor Wetenschappelijk Onderzoek" (Dutch Research Council) NWO-TTW Perspectief program P21-20 "Optical coherence; optimal delivery and positioning" (OPTIC) with TUE, TUD, ARCNL, Anteryon, ASML, Demcon, JMO, Signify, and TNO.





\bmsection{Disclosures} The authors declare no conflicts of interest.

 \end{backmatter}

\bibliography{applic_wfs}

\bibliographyfullrefs{applic_wfs}

\end{document}